# A simple setup to measure muon lifetime and electron energy spectrum of muon decay and its Monte Carlo simulation

Yushi Hu[#], Tianye Wang[#], Yefeng Mei, Zhao Zhang, and Chuangang Ning*

*Department of Physics, State Key Laboratory of Low-Dimensional Quantum Physics, Tsinghua University, Beijing 100084, China*

[#]The authors are students of The High School Affiliated to Renmin University of China, Beijing.

*Email: ningcg@tsinghua.edu.cn

## Abstract

We designed a simple setup to measure the muon lifetime and the electron energy spectra of muon decay. A low cost coincidental circuit was designed to select the signals of muon decay events detected by a plastic scintillator detector. It triggered a digital oscilloscope to record the signals of muon decay events for measuring the muon lifetime and electron energy spectrum. A Landau-distribution energy loss method was introduced to conduct the energy calibration of the system. The experimental results were well reproduced by the Monte Carlo simulation. The software and hardware of the system are completely open to students, thus more helpful for instruction and motivation.

## I. Introduction

Muon ($\mu$), the heavy cousin of electron, is a truly elementary particle. It has a unitary negative electric charge of $-1$ and a spin of 1/2, a much greater mass (105 MeV/$c^2$) [1-3]. The antiparticle of muon is $\mu^+$, with a charge state +1. On Earth, most of the naturally occurring muons are created by pions ($\pi^- \rightarrow \mu^- + \bar{\nu}_\mu$, $\pi^+ \rightarrow \mu^+ + \nu_m$) in cosmic rays. Here $\nu_\mu$ is muon type neutrino, and $\bar{\nu}_\mu$ is anti-muon type neutrino. Most of muons are created at 15 km above sea level and have strong penetrability because they do not involve in the strong interaction. At sea level about three-fourths of high energy particles are muons with an average energy of about 4 GeV. The flux of muons on sea level is about $1 \sim 2 cm^{-2} \cdot \min^{-1}$ [1-3]. Compared with the generic nuclear and particle experiments, which require radioactive sources or accelerators that are limitedly used under the safety control, muons are delivered free into the classroom by the cosmic radiation without interruption. Like most particles generated via the high energy collisions, muon is not stable. The mean lifetime of muon is 2.2 μs. The decay mode of muon is a three-body process:

$$\mu^- \rightarrow e^- + \nu_\mu + \bar{\nu}_e$$

$$\mu^+ \rightarrow e^+ + \bar{\nu}_\mu + \nu_e,$$

where $\nu_e$ and $\bar{\nu}_e$ are the electron type and anti-electron type neutrinos, respectively. Investigation of muon decay is a fascinating experiment for high school students and undergraduates [4-11]. The main obstacle is that such an experiment tends to rely on the relatively expensive and bulky electronic instrumentation (e.g., NIM and CAMAC-standard modules) and stand-alone multi-channel analyzers, which are generally not available in schools and require a background knowledge prerequisite for understanding the operation principles. Recently, Ye and coworkers designed a compact apparatus based on the specially designed readout instrument, which greatly reduced the cost [9]. As a result, more and more students have chances to conduct muon experiments. In this work, we report a simple setup based on a common digital oscilloscope to visualize and to record the signals. The data are analyzed either by a home-coded program in real time or by a program off-line. In addition to the measurement of the muon lifetime, our approach also measures the pulse height distributions of the muon signals and electron signals from the muon decay events. In order to convert the pulse height distributions into the energy spectra, a suitable calibration method is required. The main obstacle is that there is no suitable radioactive source that can emit gamma rays or charged particles with an energy around 50 MeV. In this work, we introduce the Landau-distribution [12] energy loss method. It is well known that the energy loss by muons passing through the matter is proportional to the amount of matter they traverse. The specific energy loss for muons is about 2 MeV per g/cm$^2$[13], which is not sensitive to the energy of muons. Thus, the energy loss for a specific traversing distance can be determined, which is then used to calibrate the system.

## II. Experimental methods

Now the digital oscilloscopes have become common tools in laboratories due to the greatly reduced cost. For example, the price of a digital oscilloscope with 100 MHz analog bandwidth, 1 G/s sample rate is roughly USD$1000. A digital oscilloscope usually has a USB interface for connection with a computer. With the remote control of a computer, a digital oscilloscope can be used as a data acquisition card. Compared with the general digitizer, a digital oscilloscope's advantages are the very high sample rate and the very high dynamic range of input signals varied from millivolt to 20 Volt. The disadvantage is the low data transfer rate. Typically, the host computer can only read data ~20 times per second via a USB interface. However, it is fast enough for handling the cosmic muon decay events in undergraduate physics laboratories. For example, the count rate of muon decay events in a 25 kg plastic scintillator detector is ~0.1/s, though the count rate of signals produced by the muons passing through the detector is much higher. Therefore, a coincidental circuit was specially designed to select the muon decay events. If a muon decays in the detector, we will see an electron signal immediately after the muon signal. Here, we assume that the signal indicates a muon decay event if the interval between two pulses is less than 22μs (10 times of the muon life time). Fig.1 illustrates the schematic diagram of the measurement.

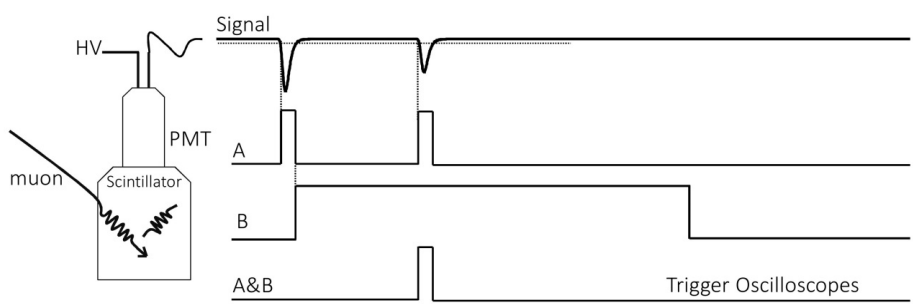

Fig.1. The schematic diagram of the selection of muon decay events. The amplitude of muon signal is compared with a threshold indicated by the horizontal dotted line. If it is larger than the threshold, a narrow pulse will be generated ( "A", typical width 50 ns), and then a wide pulse ( "B", typical width 22 μs) will be generated at the falling edge of pulse "A". If another pulse "A" is generated subsequently by the electron produced in the muon decay event during the pulse "B", a pulse "A&B" will be generated to trigger the digital oscilloscope to record the wave form of signals.

Fig.2 shows the schematic circuit made for the realization of the functions illustrated in Fig.1. A fast comparator LM710 (response time 40 ns) is used as a pulse height discriminator. The threshold voltage of the comparator can be adjusted by a potentiometer. Its output is transformed to the standard TTL pulse by a NAND gate (74LS00), and then triggers a monostable multivibrator (74LS221) via the negative-transition-triggered input (U1B, pin9). Its output, a narrow pulse with 50 ns width (U1B, pin5), triggers another monostable multivibrator (U4A) also via the negative-transition-triggered input (U4A, pin1) to generate a wide pulse with 22 μs width (U4A, Pin13). The output of the Boolean NAND operation (U2B) between the narrow pulse and the wide pulse is used as a trigger signal of a digital oscilloscope. The flash of LED D1 indicates muon signals, while D2 indicates muon decay events. This simple circuit can be built by most of undergraduate students with a little experience in electronics.

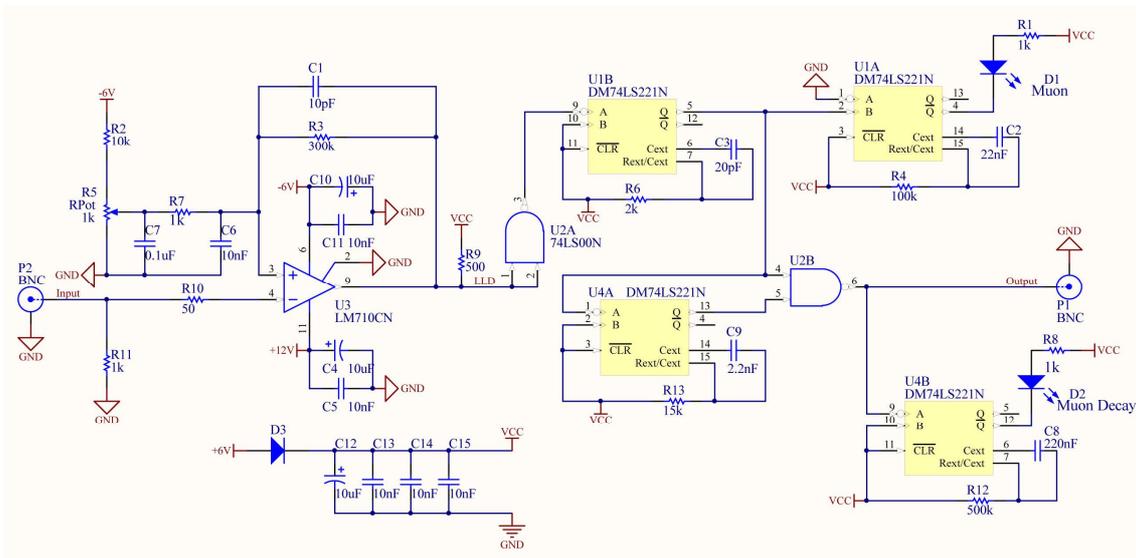

Fig.2. (Color online) The schematic circuit for the selection of muon decay events.

The heart of the detector is a cylinder of 31 cm diameter and 35 cm length made of plastic scintillating material called NE 102. Its density is 1.02 g/cm$^3$. When a high energy muon or electron passes through the scintillator, tiny light flashes are produced along its track. These flashes will be detected by a photomultiplier tube (PMT). The surfaces of scintillator are coated with ZrO2 reflection layer for enhancement of light collection, and then wrapped carefully with black paper to prevent light leaks. The PMT we used is GDB52B type, which is harvested from an old NaI(Tl) gamma detector. In present experiment, the PMT works at −1200 V provided by an adjustable high voltage (HV) power supply. The PMT has a preamplifier integrated in the shield tube. The typical amplitude of its output is ~0.5 volt, with a pulse width 20 ns. In a typical muon decay event, the first pulse is produced by the muon entering the detector, and the second pulse is produced by the electron generated by the muon decay. The time interval between the two pulses is the lifetime of the muon. The amplitude of the signal is proportional to the energy loss of the charged particle deposited in the scintillator. Therefore, the measurement of the time intervals between the two signals produces the lifetime spectrum. The measurement of amplitudes of the signals produces the energy spectra. In order to calibrate the energy scale of the system, we use two small scintillation detectors to guarantee the cosmic muons vertically pass through the detector. The specific energy loss for muons is about 2 MeV per $g/cm^2$. The density of the plastic scintillating material is 1.02 $g/cm^3$. Therefore, the specific energy loss is about 2 MeV/cm in the main detector. The size of the scintillator of the small detector is 34 cm × 34 cm square with a thickness 2.5 cm. See Fig.3 for the layout of the three detectors. One small detector locates at the position 88 cm above the main detector and the other is 112 cm below the main detector. This limits the incident angle of muons in the range of ±9.6°. A fast coincidence module with a 100 ns window is used to select muons that vertically pass through the three detectors. The coincident output is used to trigger the oscilloscope for recording the signals of the main detector. As shown later, the energy loss in the main detector has a Landau distribution.

## III. Experimental results and Monte Carlo simulation

A program written by using Visual Basic language was designed to record the wave forms of the signals, to analyze the time intervals and amplitudes of signals of decay events, to display and save the lifetime spectrum, energy spectra of muons and electrons. The interface of the program has four panels for the real time display of the waveform of muon decay signals: the statistical distribution of muons lifetimes, the statistical distribution of muon signal amplitudes, and the statistical distribution of electron signal amplitudes. Once the digital oscilloscope is triggered, it will sample the signals for duration of 40 μs and store it in a hard disk. The 40 μs duration includes 10μs before the triggering and 30μs after the triggering. The program automatically measures the time intervals and the amplitudes of the pulses, and then updates the statistics. In order to acquire all possible muon decay signals, the threshold voltage of the pulse-height discriminator is usually set just above the noise level. Since all the waveforms of signals are saved by the computer each by each, the offline analysis can set new criteria to choose the valid muon decay events more strictly.

**Energy Calibration.** Fig.3 shows the Landau-distribution energy loss of muons that pass through the main detector. Generally, the pulse height is proportional to the energy loss deposited in the detector. In order to obtain the conversion parameter $k$ (unit MeV/mV) between the energy loss and the pulse height, we use the GEANT 4 [14] program to simulate the process. The energy distribution of cosmic muons is obtained from Ref.[15], and the incidental angle $q$ distribution is according to the angular distribution $\cos^2 q$. The energy loss in the main detector is recorded if the muon passes through the three detectors. Fig.3 b) is the simulated results. It can be seen that the peak position of the Landau distribution located at 71 MeV, which agree well with the estimation 35 cm × 2 MeV/cm. Fig.3 c) is the measured pulse height distribution. The measured result is also a Landau-type distribution. The measured distribution is broader than the simulated one, which might be due to the non-perfect photon collection of our detector. By comparing the experimental distribution with the simulated one, we can obtain the conversion factor $k$=0.215 MeV/mV. It should be noted that $k$ depends on the high voltage of PMT and the gain of amplifiers.

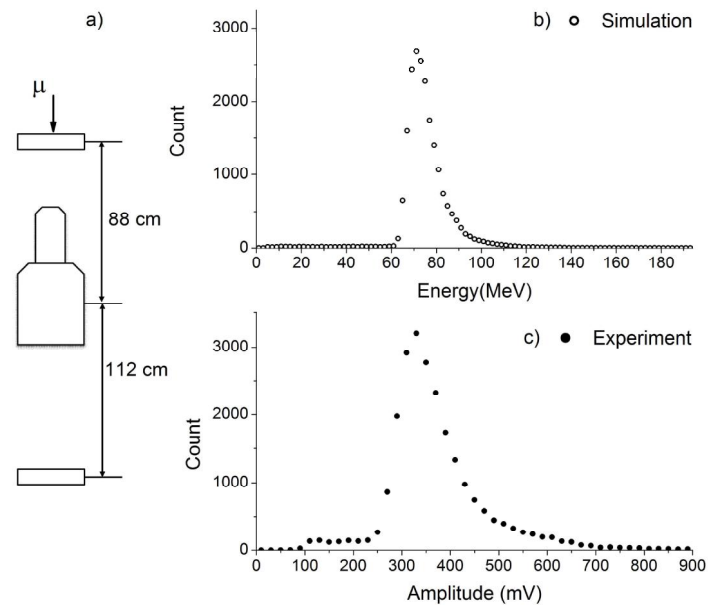

Fig.3 Landau distributions of energy loss spectra of muons. a) The layout of the three detectors for selecting the muons those vertically pass through the main detector. b) The Monte Carlo simulated results. c) The experimental results.

**Muon lifetime.** Like the lifetime of radioactive particles, the lifetime of muons is a statistic averaged value on a large number of decay events. For an individual muon, the decay happens randomly, i.e. when it will happen cannot be predicted exactly. To measure the mean lifetime of muons, we have collected $1.1\times10^5$ muon decay events for one week. A decay-time histogram is shown in Fig. 4 c). Dots are experimental data, and the red line is an exponential fitting of the data using $N(t)=N_0+A_0\mathrm{Exp}(-t/t)$. Here $t$ is the time interval between the muon signal and the electron signal, $t$ is the mean lifetime of muons, $N_0$ is a background due to accidental coincidence, and $A_0 t$ is the total number of muons. We observed an interesting phenomenon during the experimental period. At the beginning, we observed a hump at the decay time $t\approx 7$ μs in the lifetime spectrum. The hump is due to pseudo-events caused by the ion feedback of photomultiplier, as called after-pulse problem. However, the hump gradually disappeared after the PMT was applied with the high voltage for several months. The after-pulse problem is due to the residual gases inside a photomultiplier tube, which can be ionized by the collision with electrons. When these ions strike the photocathode or earlier stages of dynodes, secondary electrons may be emitted, thus resulting in relatively large output noise pulses. These noise pulses are usually observed as after-pulses following the primary signal pulses. A photomultiplier with special design to suppress the ion feedback, or using two photomultipliers solution can solve this problem [8]. In our case, one possible explanation is that the residual gases were gradually gotten rid of due to the long time electron impact ionization. The mean lifetime for mixed $\mu^-$ and $\mu^+$ we obtained is $t=2.098\pm0.006$ μs (statistic error only), which is less than the free space value $t_\mu=2.196\,9811 \pm 0.000\,0022$ μs [15, 17]. This is due to the non-negligible probability that a $\mu^-$ will be captured into the K-shell of a scintillator carbon atom and then be absorbed by its positively charged nucleus without producing a high energy decay electron [9, 16].

**Energy spectra.** In addition to the muon lifetime, the electron spectrum from the muon decay is also of great theoretical interest [18]. The amplitudes of signals are proportional to the energy of the charged particle if it loses all energy in the scintillator. Therefore, if the detector size is large enough, we should be able to observe the energy spectrum of electrons from the muon decay. Fig.4 b) and d) show the energy spectrum of muons that stop in the detector and the energy spectrum of electrons that were produced by the decay event, respectively. The conversion factor

$k$=0.215 MeV/mV has been used. At first sight, the shape of the energy distribution of electrons is different from that observed by the advanced experimental techniques [17]. The intensity in the low energy region is higher than the prediction by the standard model [17]. This is due to that the size of our detector is not large enough. Generally, high-energy particles will not be fully stopped in the plastic scintillator; they will pass through, leaving a part of their energy in it. The amount of this energy loss ($\Delta E$) can be estimated from "energy-loss tables" given in literature [19]. The maximum energy of electron from the muon decay is 52.5 MeV [18]. The electron range, a measure of the straight-line penetration distance of electrons in a solid, is 20 cm for $E$=52.5 MeV in the plastic scintillator [19]. Obviously, the size of our detector $\Phi$31 cm×35 cm is not large enough to fully stop all electrons. For cosmic muons, a specific energy-loss value of $dE/dx \approx$ 2 MeV/cm yields in our plastic cylinder a maximum energy loss of about $\Delta E$ = 70 MeV for vertically travelling. This can explain why amplitudes of muon signals are usually larger than that of electron signals in Fig.4.

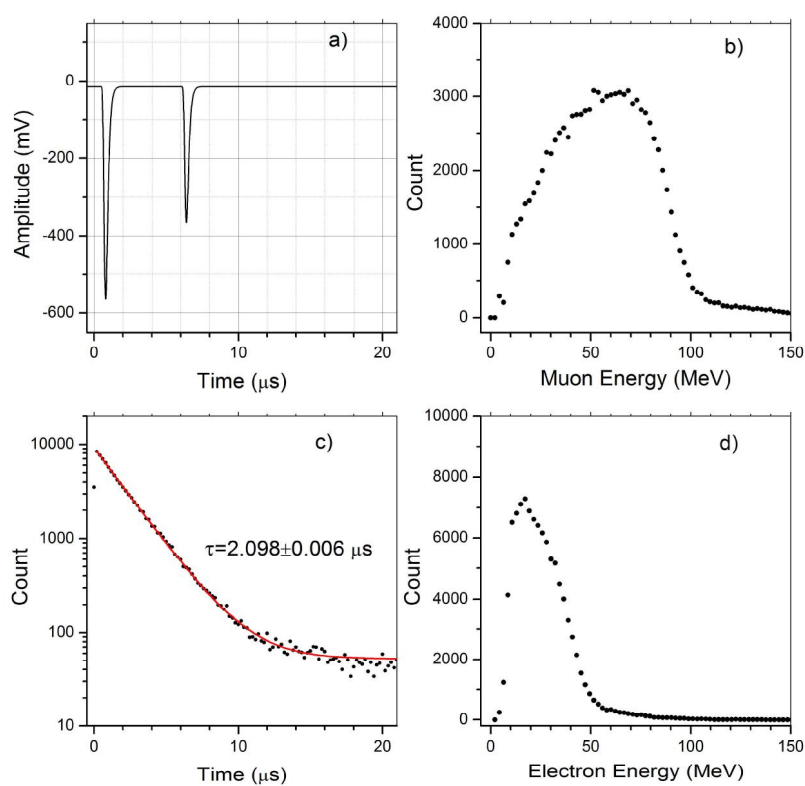

Fig.4 (Color online) The program interface for real time display of a) the waveform of muon decay signal just captured; b) the statistical distribution of muon signal amplitudes; c) the statistical distribution of lifetimes of muons. The line is the fitting curve to the function $N(t)=N_0+A_0\mathrm{Exp}(-t/t)$; d) the statistical distribution of electron signal amplitudes.

In order to explain the observed energy spectrum of electrons produced in the muon decay events, we simulated the process using GEANT 4 program [14]. Without loss of generality, we assume muons stop in the detector, and the positions of muons are uniformly distributed in the detector. As shown in Fig.5, the shape of the simulated result agrees well with the experimental one. The simulation has a clear cut at 52.5 MeV, which is consistent with the theoretical maximum energy 52.5 MeV of electrons produced by the muon decay. The energy spectrum observed in Fig.5 can rule out the two-body decay, because a two-body decay will produce mono-energetic electrons. The energy is $\frac{1}{2}m_\mu c^2$, 52.5 MeV.

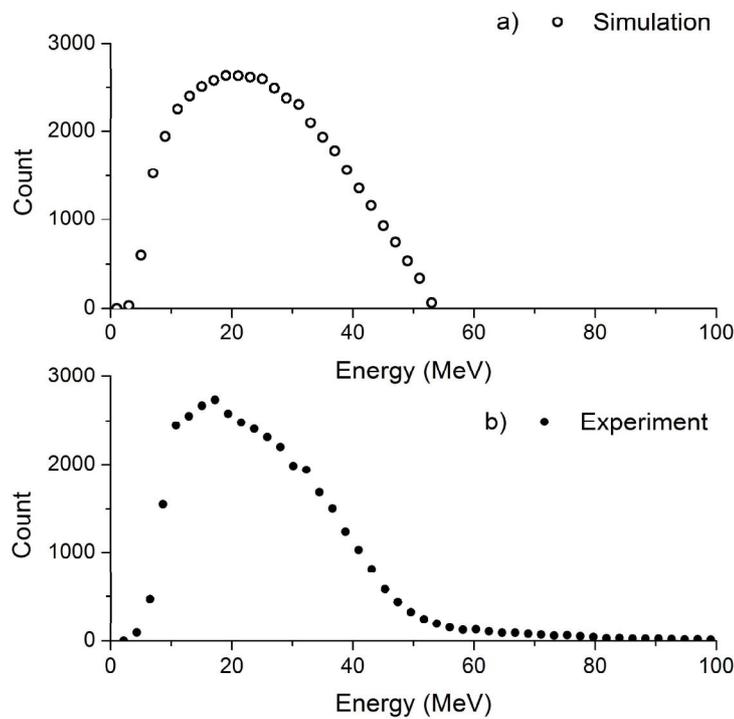

Fig.5 The energy spectrum of electrons produced in the muon decay events. a) The Monte Carlo simulation. b) The experimental results.

## IV. Summary

We have designed a simple setup for the muon decay measurement. A simple circuit for the selection of muon decay events was designed to trigger a digital oscilloscope for recording the signals. To overcome the difficulty of the detector energy calibration around energy 50 MeV, a Landau-distribution energy loss method was introduced. The measurements of muon lifetime and electron energy spectrum from the muon decay were demonstrated. The measured results were well reproduced by the Monte Carlo simulation. The system is simple and low cost, and the signals processing is transparent. Therefore, it is more helpful for instruction and motivation to students. The idea using a coincidental circuit to trigger a digital oscilloscope for recording the signals, as demonstrated here, is also applicable to other coincidental measurement if the coincidental count rate is not high. The copy of the circuit and codes can be obtained free upon request by sending Email to the author NCG.